\documentclass[prl,aps,epsfig,superscriptaddress,showpacs]{revtex4}
\usepackage{bm}
\usepackage{amsfonts}
\usepackage[dvips]{graphicx}
\usepackage{mathrsfs}
\usepackage[intlimits]{amsmath}
\usepackage[colorlinks, citecolor=red]{hyperref}

\begin{document}

\title{Quantum teleportation from light beams to vibrational states \\
of a macroscopic diamond}
\author{P.-Y. Hou$^{1}$, Y.-Y. Huang$^{1}$, X.-X. Yuan$^{1}$, X.-Y. Chang$%
^{1}$, C. Zu$^{1}$, L. He$^{1}$, L.-M. Duan}
\affiliation{Center for Quantum Information, IIIS, Tsinghua University, Beijing 100084,
PR China}
\affiliation{Department of Physics, University of Michigan, Ann Arbor, Michigan 48109, USA}

\begin{abstract}
With the recent development of optomechanics, the vibration in solids,
involving collective motion of trillions of atoms, gradually enters into the
realm of quantum control. Built on the recent remarkable progress in optical
control of motional states of diamonds, here we report an experimental
demonstration of quantum teleportation from light beams to vibrational
states of a macroscopic diamond under ambient conditions. Through quantum
process tomography, we demonstrate average teleportation fidelity $(90.6\pm
1.0)\%$, clearly exceeding the classical limit of $2/3$. The experiment
pushes the target of quantum teleportation to the biggest object so far,
with interesting implications for optomechanical quantum control and quantum
information science.
\end{abstract}

\maketitle



\section{Introduction}

\textbf{\ }Quantum teleportation has found important applications for
realization of various quantum technologies \cite{1,2,3,4}. Teleportation of
quantum states has been demonstrated between light beams \cite{6,7,8,8b},
trapped atoms \cite{9,10,11,12}, superconducting qubits \cite{13}, defect
spins in solids \cite{14}, and from light beams to atoms \cite{15,16} or
solid state spin qubits \cite{17,18}. It is of both fundamental interest and
practical importance to push quantum teleportation towards more macroscopic
objects.\textbf{\ }Observing quantum phenomenon in macroscopic objects is a
big challenge as their strong coupling to the environment causes fast
decoherence which quickly pushes them to the classical world. For example,
quantum coherence is hard to survive in mechanical vibration of macroscopic
solids, which involves collective motion of a large number of strongly
interacting atoms. Despite this challenge, achieving quantum control for the
optomechanical systems becomes a recent focus of interest with remarkable
progress \textbf{\textbf{\cite{19,20,21,22,22a1,22a2,23,24,24a,25,26,27}}}.
This is driven in part by the fundamental interest and in part by the
potential applications of these systems for quantum signal transduction
\textbf{\textbf{\cite{23,24,24a}}}, sensing \textbf{\textbf{\cite{19}}}, and
quantum information processing \textbf{\textbf{\cite{19,20,21}}}. There are
typically two routes to achieve quantum control for the optomechanical
systems: one needs to either identify some isolated degrees of freedom in
mechanical vibrations and cool them to very low temperature to minimize
their environmental coupling \textbf{\textbf{\cite{19,25,26,27}}}, or use
the ultrafast laser technology to fast process and detect quantum coherence
in such systems \textbf{\textbf{\cite{20,21,22,22a1,22a2}}}. A remarkable
example for the latter approach is provided by the optomechanical control in
macroscopic diamond samples \textbf{\textbf{\cite{20,21}}}, where the
motions of two separated diamonds have been cast into a quantum entangled
state \textbf{\textbf{\cite{20}}}.

In this paper, we report an experimental demonstration of quantum
teleportation from light beams to the vibrational states of a macroscopic
diamond sample of $3\times 3\times 0.3$ mm$^{3}$ in size under ambient
conditions. The vibration states are carried by two optical phonon modes,
representing collective oscillation of over $10^{16}$ carbon atoms. To
facilitate convenient qubit operations, we use the dual-rail representation
of qubits instead of the single-rail encoding used in previous experiments
\textbf{\textbf{\cite{20,21,22,22a1}}} and generate entanglement between the
paths of a photon and different oscillation patterns of the diamond
represented by two phononic modes. Using quantum state tomography, we
demonstrate entanglement fidelity of $(81.0\pm 1.8)\%$ with the raw data and
of $(89.7\pm 1.2)\%$ after the background noise subtraction. Using this
entanglement, we prepare arbitrary polarization states for the photon and
teleport these polarization states to the phonon modes with the Bell
measurements on the polarization and the path qubits carried by the same
photon. The teleportation is verified by quantum process tomography, and we
achieve a high average teleportation fidelity, about $(90.6\pm 1.0)\%$ (or $%
(82.9\pm 0.8)\%$) after (or before) subtraction of the background noise. To
verify the phonon's state before its fast decay, our implementation of
teleportation adopted the technique of reversed time ordering introduced in
Ref. \textbf{\textbf{\cite{20}}} where the phonon's state is read out before
the teleportation is completed. Similar to the pioneering teleportation
experiment of photons \cite{6}, our implementation of teleportation is
conditional as the Bell measurements are not deterministic and require
postselecting of successful measurement outcomes.

\section{Results}


\begin{figure}[tbp]
\includegraphics[width=8.5cm,height=7cm]{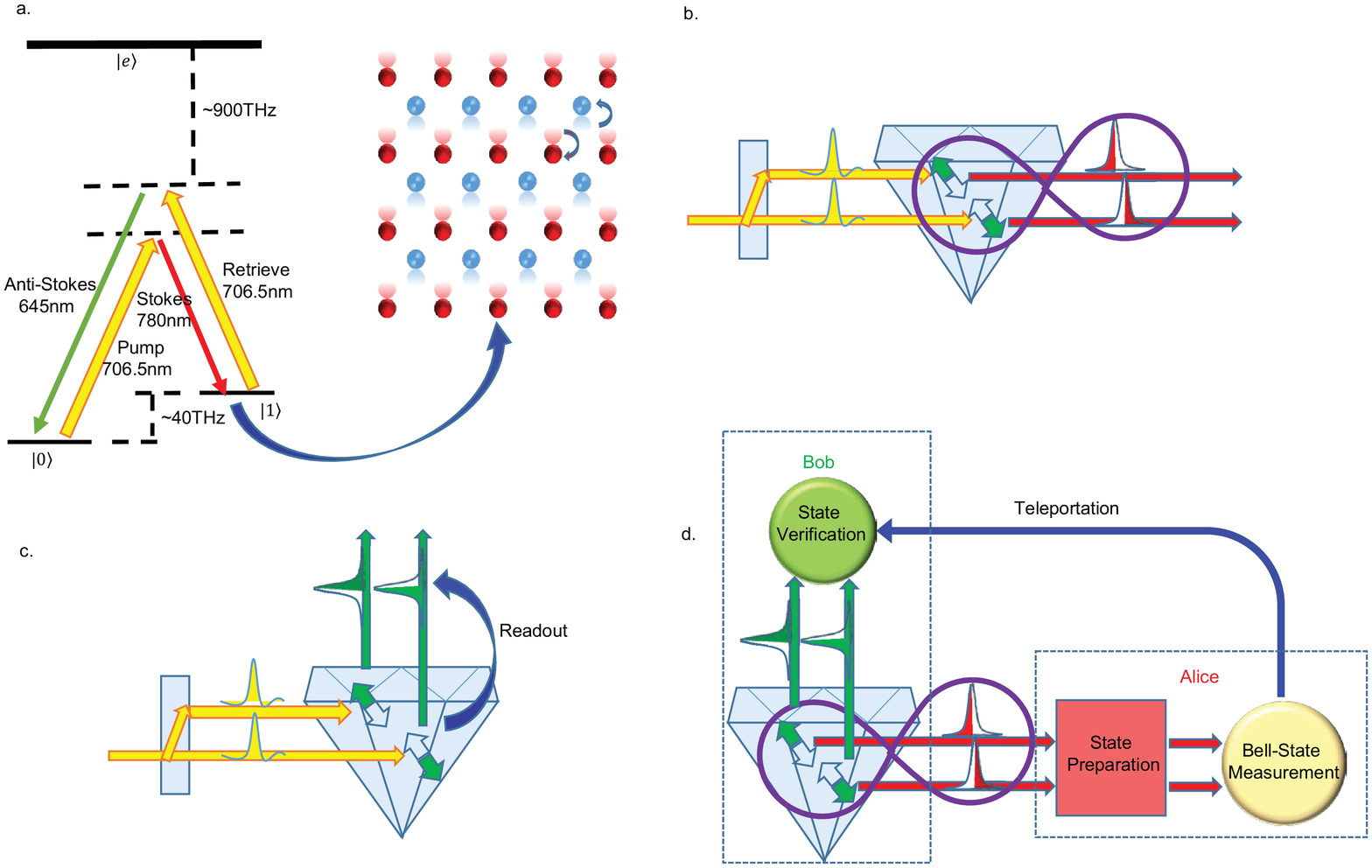}
\caption[Fig. 1 ]{\textbf{A scheme for entanglement generation and quantum
teleportation between light beams and vibrational modes of a diamond.}
\textbf{a}, Illustration of the relevant level structure in the diamond. A
write beam pumps the diamond in the ground state $\left\vert 0\right\rangle$
and generates a Stokes photon in the forward direction and an excitation in
the optical phonon mode of the diamond (denoted by the state $\left\vert
1\right\rangle$). The optical phonon mode corresponds to relative
oscillation of the atoms in each unit cell of the diamond lattice, as
illustrated by the figure on the right side. A read beam after a
controllable delay converts the phonon excitation to an anti-Stokes photon
which can be used for state readout. The corresponding wavelengths and
frequencies are shown in the figure. The state $\left\vert e\right\rangle$
denotes the electron conduction band which is far detuned from the optical
excitation. \textbf{b}, A scheme for generation of entanglement between a
phonon in the diamond and a propagating photon. The phonon state is
represented by a superposition of different oscillation modes of the
diamond, while the photon state is represented by its spatial modes. \textbf{%
c}, Readout of the phonon state with the read beams by coherently converting
the phonon modes into the corresponding anti-Stoke photon modes. \textbf{d},
A teleportation scheme using the photon-phonon entanglement. An input state
is prepared by the message sender, Alice, on the photon's polarization
degree of freedom. The photon thus carries two qubits, one by its
polarization and one by its spatial modes. Alice performs Bell measurements
on these two qubits. Conditional on certain measurement outcomes, the phonon
state is projected to the same state input on Alice's side, which is read
out and verified by Bob, the message receiver. }
\end{figure}

\subsection{Photon-to-phonon teleportation scheme}

We illustrate our entanglement generation and quantum teleportation scheme
in Fig. 1, using a type IIa single-crystal synthetic diamond sample cut
along the $100$ face from the Element Six company. Due to the strong
interaction of atoms in the diamond, the optical phonon mode, which
represents relative oscillation of the two sublattices in the stiff diamond
lattice (see Fig. 1a), has a very high excitation frequency about $40$ $THz$
near the momentum zero point in the Brillouin zone. The corresponding energy
scale for this excitation is significantly higher than the room temperature
thermal energy (about $6$ $THz$), and thus the optical phonon mode naturally
stays at the vacuum state under ambient conditions, which simplifies its
quantum control \textbf{\textbf{\cite{20,21}}}. The coherence life time of
the optical phonon mode is about $7$ ps at room temperature, which is short
but accessible with the ultrafast laser technology for which the operational
speed can be up to about $10$ THz \textbf{\textbf{\cite{20,21}}}.

We excite the optical phonon modes through ultrafast laser pulses of
duration around $150$ fs from the Ti-Sapphire laser, with the carrier
wavelength at $706.5$ nm. The diamond has a large bandgap of $5.5$ ev, so
the laser pulses are far detuned from the conduction band with a large gap
about $900$ THz. Each laser pulse generates, with a small probability $p_{%
\text{s}}$, an excitation in the optical phonon mode and a Stokes photon of
wavelength $780$ nm in the forward direction (see Fig. 1a). The relevant
output state has the form

\begin{equation}
|\Psi \rangle =\left[ 1+\sqrt{p_{\text{s}}}b_{\text{n}}^{\dag }a_{\text{t}%
}^{\dag }+o\left( p_{s}\right) \right] |\text{vac}\rangle ,
\end{equation}%
where $b_{\text{n}}^{\dag }$ and $a_{\text{t}}^{\dag }$ represent,
respectively, the creation operators for an optical phonon and a Stokes
photon, and $|$vac$\rangle $ denotes the common vacuum state for both the
photon and the phonon modes.

To generate entanglement, we split the laser pulse into two coherent paths
as shown in Fig. 1b, and the pulse in each path generates the corresponding
phonon-photon correlated state described by Eq. (1). When there is an output
photon, in one of the two paths, it is in the following maximally entangled
state with the phonon excitation%
\begin{equation}
|\Psi _{\text{nt}}\rangle =\left( \left\vert U\right\rangle _{\text{n}%
}\left\vert U\right\rangle _{\text{t}}+\left\vert L\right\rangle _{\text{n}%
}\left\vert L\right\rangle _{\text{t}}\right) /\sqrt{2}.
\end{equation}%
Here, $\left\vert U\right\rangle $ or $\left\vert L\right\rangle $
represents an excitation in the upper or lower path, and its subscript
denotes the nature of the excitation, "n" for a phonon and "t" for a photon.
We drop the vacuum term in Eq. (1) as it is eliminated if we detect a photon
emerging from one of the two paths. After entanglement generation, the
photon state can be directly measured through single-photon detectors. To
read out the phonon state, we apply another ultrafast laser pulse after a
controllable delay within the coherence time of the optical phonon mode and
convert the phononic state to the same photonic state in the forward
anti-Stokes mode at the wavelength of $645$ nm (see Fig. 1c). The state of
the anti-Stokes photon is then measured through single-photon detectors
together with linear optics devices. Note that the retrieval laser pulse
could have a carrier frequency $\omega _{\text{r}}$ different from that of
the pump laser. For instance, with $\omega _{\text{r}}$ near the telecom
band, our teleportation protocol would naturally realize a quantum frequency
transducer that transfers the photon's frequency to a desired band without
changing its quantum state. A quantum frequency transducer is widely
recognized as an important component for realization of long-distance
quantum networks \textbf{\textbf{\cite{23,24,24a}}}.

To realize teleportation, we need to prepare another qubit, whose state will
be teleported to the phonon modes in the diamond. Similar to the
teleportation experiments in Refs. \textbf{\textbf{\cite{7,16}}}, we use the
polarization state of the photon to represent the input qubit, which can be
independently prepared into an arbitrary state $c_{0}\left\vert
H\right\rangle _{\text{t}}+c_{1}\left\vert V\right\rangle _{\text{t}}$,
where $\left\vert H\right\rangle _{\text{t}}$ and $\left\vert V\right\rangle
_{\text{t}}$ denote the horizontal and the vertical polarization states and $%
c_{0},c_{1}$ are arbitrary coefficients. The Bell measurements on the
polarization and the path qubits carried by the same photon can be
implemented through linear optics devices together with single-photon
detection (see Fig. 1d), and the teleported state to the phononic modes is
retrieved and detected through its conversion to the anti-Stokes photon.
Same as Ref. \textbf{\textbf{\cite{20}}}, the short life time of the
diamond's vibration modes requires us to retrieve and detect the phonon's
state before applying detection on the Stokes photon, thus the phonon's
state is measured before the teleportation protocol is completed. The
reversed time ordering in this demonstration of quantum teleportation makes
it unsuitable for application in quantum repeaters which requires a much
longer memory time, however, it does not affect application of our
teleportation experiment for realization of a quantum frequency transducer
or a new source of entangled photons as discussed above.


\begin{figure}[tbp]
\includegraphics[width=17cm,height=7cm]{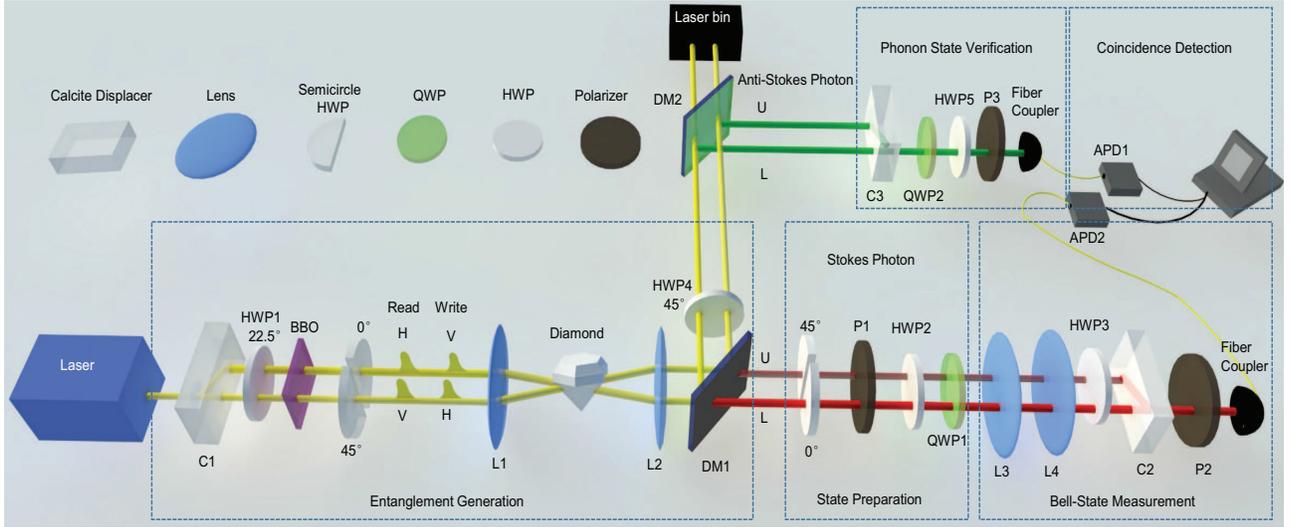}
\caption[Fig. 1 ]{\textbf{Experimental setup for entanglement verification
and quantum teleportation}. Femtosecond laser pulses from a Ti-Sapphire
laser (Coherent, driven by Verdi 18), with a repetition frequency of $76$
MHz, a carrier wavelength of $706.5$ nm, and a polarization along the $%
\left\vert H\right\rangle +\left\vert V\right\rangle $ direction, are split
by a birefringent calcite into two coherent paths with equal amplitudes.
After rotation of the pulse polarization to equal superposition of $%
\left\vert H\right\rangle $ and $\left\vert V\right\rangle $ again with a
half wave plate (HWP1) set at $22.5^{o}$, we introduce a time delay of $388$
fs to the two polarization components $H$ and $V$ with a birefringent BBO\
crystal. We use the lead pulse of $H$\ polarization as the write beam and
the lagged pulse of $V$\ polarization as the read beam. After semicircle
HWPs set at $0^{o}$ and $45^{o}$, respectively, at the upper and lower
paths, the polarization states of the pump beams are shown in the figure
before the diamond sample. The write beam is focused by the lens L1 on the
diamond sample and generates a Stokes photon in one of the paths and an
excitation in the corresponding optical phonon modes of the diamond. The
Stokes photon, at the wavelength of $780$ nm, is transmitted by the
dichromatic mirror DM1 after the collection lens L2, with its two paths
recombined by the calcite C2. To verify entanglement, we do not need HWP3
and the optical elements in the state preparation box. The lens L3 and L4
are used to adjust the distance between the two optical paths so that they
can be combined at the calcite C2. The single-photon detector APD2, together
with rotation of the polarizer P2, detects the two path (or polarization)
components of the Stokes photon in different bases. To read out the state of
the phonon modes, the read pulse converts the phonon to the anti-Stokes
photon in the corresponding paths, preserving its quantum state. The
anti-Stokes photon, at a shorter wavelength of $645$ nm, is reflected by
both of the dichromatic mirrors DM1 and DM2, with its two paths recombined
through the calcite C3. The single-photon detector APD1, together with
rotation of the half and the quarter wave plates HWP5 and QWP2, detects the
anti-Stokes photon (and thus the phonon) in any superposition bases. For
quantum teleportation, semicircle HWPs and the polarizer P1 in the state
preparation box transforms the photon-phonon entangled state to the standard
form of Eq. (2) and initializes the photon polarization to the state $%
\left\vert V\right\rangle $. The waveplates HWP2 and QWP1 then prepare the
to-be-teleported photon polarization to any superposition state. The calcite
C2, the HWP3, the polarizer P2, and the detector APD2, together, make a Bell
measurement on the two qubits carried by the polarization and the path
degrees of freedom of the same photon. The photon coincidence counts are
registered through a FPGA (Field-Programmable Gate Array) board with a $5$
nm coincidence window. }
\end{figure}

\subsection{Experimental realization of teleportation}


Our experimental setup is shown in Fig. 2. First, we verify entanglement
generated between the Stokes photon and the optical phonon modes in the
diamond. For this step, we remove the optical elements in the state
preparation box shown in Fig. 2 and set the angle of HWP3 to $0^{o}$.
Different from the scheme illustrated in Fig. 1, we insert semicircle HWPs
set at $0^{o}$ and $45^{o}$, respectively, at the upper and the lower paths
of the pump beam, so that both the Stokes photon and the anti-Stokes photon
after the retrieval pulse have orthogonal polarizations along the two output
paths, which can be combined together through the calcites C2 and C3. This
facilitates the entanglement measurement through detection in complementary
local bases by rotating the polarizers P2 and P3 and the wave plates HWP5
and QWP2. Due to the different incident directions of the pump pulses at the
upper and the lower paths, the corresponding phonon modes excited in the
diamond have different momenta, so they represent independent modes even if
they have partial spatial overlap. The phonon is converted to the
anti-Stokes photon by the retrieval pulse, so we measure the photon-phonon
state by detecting the coincidence counts between Stokes and anti-Stokes
photons in different bases. In Fig. 3a, we show the registered coincidence
counts as we rotate the angle of the polarizer P2. The oscillation of the
coincidence counts with a visibility of $(74.6\pm 3.6)\%$ is an indicator of
coherence of the underlying state. To verify entanglement of the
photon-phonon state, we use quantum state tomography to reconstruct the full
density matrix from the measured coincidence counts \textbf{\textbf{\cite{28}%
}}, with the resulting matrix elements shown in Fig. 3b. From the
reconstructed density matrix $\rho _{\text{e}}$, we find its entanglement
fidelity, defined as the maximum overlap of $\rho _{\text{e}}$ with a
maximally entangled state, $F_{\text{e}}=(81.0\pm 1.8)\%$, significantly
higher than the criterion of $F_{\text{e}}=0.5$ for verification of
entanglement \textbf{\textbf{\cite{29}}}. The error bars are determined by
assuming a Poissonian distribution for the photon counts and propagated from
the raw data to the calculated quantities through exact numerical
simulation. The dominant noise in this system comes from the accidental
coincidence between the detected Stokes and the anti-Stokes photons \textbf{%
\textbf{\cite{20,21}}}. To measure the contribution of this accidental
coincidence, we introduce an extra time delay of $13$ $ns$, the repetition
period of our pump pulses, to one of the detectors when we record the
coincidence. When we subtract the background noise due to this accidental
coincidence, the resulting matrix elements reconstructed from the quantum
state tomography are shown in Fig. 3c. We find the entanglement fidelity is
improved to $F_{\text{e}}=(89.7\pm 1.2)\%$ after subtraction of the
accidental coincidence.

\begin{figure}[tbp]
\includegraphics[width=8.5cm,height=7cm]{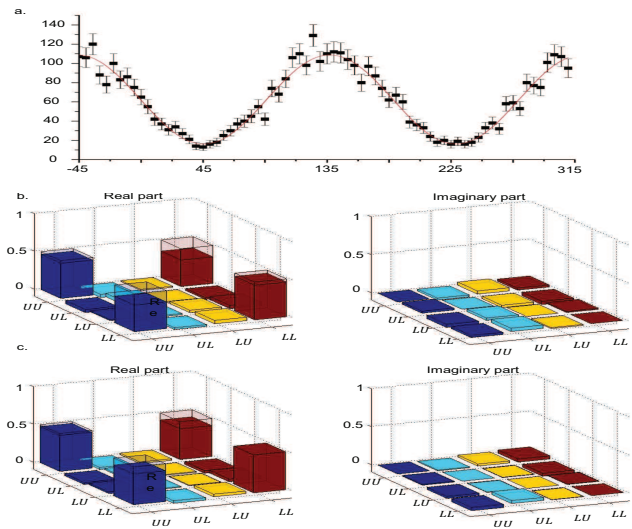}
\caption[Fig. 3 ]{\textbf{Verification of photon-phonon entanglement between
light beams and the vibrational modes of the diamond.} \textbf{a}
Coincidence counts of Stokes and anti-Stokes photons as a function of the
rotation angle (in degree) of the polarizer (P2 in Fig. 2) for the Stokes photon when
the measurement basis of the anti-Stokes photon is fixed at $\left\vert
U\right\rangle -\left\vert L\right\rangle $. The error bars denote the
standard deviation. \textbf{b}, Real and imaginary parts of the density
matrix elements for the phonon-photon entangled state reconstructed through
the quantum state tomography. The hollow caps correspond to the values of
matrix elements for a perfect maximally entangled state. \textbf{c}, Same as
Fig. 3b, but we subtract the background noise due to the accidental
coincidences of the photon detectors. The coincidence count rate for Stokes
and anti-Stokes photons is $8$ per second for measurements
in the $UU$ and $LL$ bases. }
\end{figure}

To perform quantum teleportation using the photon-phonon entanglement, we
first transform the effective photon-phonon entangled state to the standard
form of Eq. (2) by the semicircle HWPs in the state preparation box of Fig.
2. The polarizer P1 and the waveplates HWP2 and QWP1 then prepare the
to-be-teleported photon polarization to arbitrary superposition states $%
|\Phi _{\text{in}}\rangle =c_{0}\left\vert H\right\rangle _{\text{t}%
}+c_{1}\left\vert V\right\rangle _{\text{t}}$. We perform Bell measurement
through the calcite C2, the HWP3, the polarizer P2, and the detector APD2.
For instance, with the HWP3 set at $0^{o}$ and the polarizer P2 set along
the direction $\left\vert H\right\rangle +\left\vert V\right\rangle $, a
photon count in the detector APD2 corresponds to a projection to the Bell
state $\left( \left\vert H\right\rangle _{\text{t}}\left\vert U\right\rangle
_{\text{t}}+\left\vert V\right\rangle _{\text{t}}\left\vert L\right\rangle _{%
\text{t}}\right) /\sqrt{2}$ for the polarization and the path qubits of the
photon before the measurement box. By rotating the angles of HWP3 and P2, we
can also perform projection to any other Bell states.

The experimental result for teleportation is shown in Fig. 4. The
teleportation fidelity is defined as $F=\left\langle \Phi _{\text{in}%
}\right\vert \rho _{\text{out}}|\Phi _{\text{in}}\rangle $, where $|\Phi _{%
\text{in}}\rangle $ is the input state at Alice's side and $\rho _{\text{out}%
}$ denotes the output density matrix at Bob's side reconstructed through
quantum state tomography measurements. In Fig. 4a, we show the teleportation
fidelity under six complementary bases states with $|\Phi _{\text{in}%
}\rangle =\left\vert H\right\rangle _{\text{t}},\left\vert V\right\rangle _{%
\text{t}},\left\vert \pm \right\rangle _{\text{t}}=\left( \left\vert
H\right\rangle _{\text{t}}\pm \left\vert V\right\rangle _{\text{t}}\right) /%
\sqrt{2},$ $\left\vert L\right\rangle _{\text{t}}=\left( \left\vert
H\right\rangle _{\text{t}}\pm i\left\vert V\right\rangle _{\text{t}}\right) /%
\sqrt{2},\left\vert R\right\rangle _{\text{t}}=\left( \left\vert
H\right\rangle _{\text{t}}-i\left\vert V\right\rangle _{\text{t}}\right) /%
\sqrt{2}$ in cases with and without subtraction of the background noise. The
average fidelity over these six bases states is $(93.9\pm 0.8)\%$ (or $%
(83.0\pm 0.8)\%$) with (or without) background noise subtraction. This
average fidelity is significantly higher than $2/3$, the boundary value for
the fidelity that separates quantum teleportation from classical operations.
For more complete characterization, we also perform quantum process
tomography (QPT) for the teleportation operation. In the ideal case,
teleportation should be characterized by an identity transformation, meaning
that Alice's input state is teleported perfectly to Bob's side. The
experimentally reconstructed process matrix elements are shown in Fig. 4b
(see Methods for explanation of QPT). The process fidelity is given by $F_{%
\text{p}}=(85.9\pm 1.6)\%$, which corresponds to a teleportation fidelity $%
\overline{F}=(90.6\pm 1.0)\%$ averaged over all possible input states with
equal weight in the qubit space.

\begin{figure}[tbp]
\includegraphics[width=8.5cm,height=7cm]{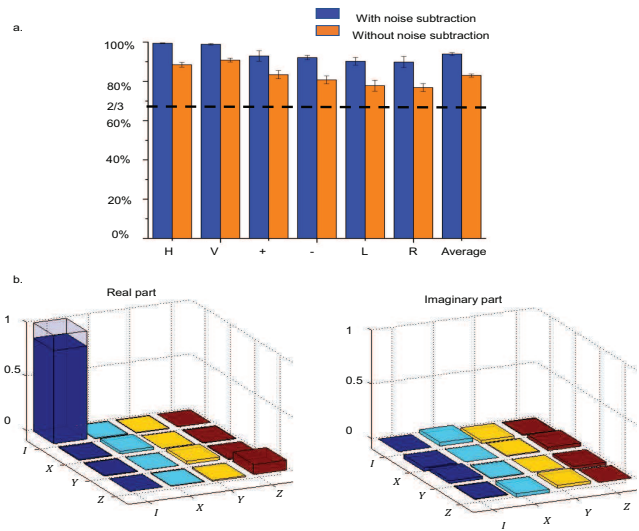}
\caption[Fig. 3 ]{\textbf{Experimental results for teleportation of quantum
states from light beams to the diamond vibrational modes.} \textbf{a}, The
teleportation fidelities for the six complementary basis states. The last
two columns show the teleportation fidelity averaged over these six input
states. The results are shown for both cases with or without subtraction of
the background noise. The error bar denotes one standard deviation. The
dashed line at fidelity $2/3$ corresponds to the classical-quantum boundary
for teleportation. \textbf{b}, Real and imaginary parts of the process
matrix elements for quantum teleportation reconstructed through the quantum
process tomography (see Methods). The hollow caps correspond to the values
of process matrix elements for a perfect teleportation operation. }
\end{figure}

\section{Discussion}

Teleportation of quantum states from a photon to the vibration modes of a
millimeter-sized diamond under ambient conditions generates a quantum link
between the microscopic particle and the macroscopic world around us usually
under the law of classical physics. In our experiment, the ultrafast laser
technology provides the key tool for fast processing and detection of
quantum states within its short life time in macroscopic objects consisting
of many strongly interacting atoms that are coupled to the environment.
Combined with the tunability of the wavelength for the retrieval laser pulse
\textbf{\textbf{\cite{22a1}}}, the technique introduced in our experiment
would be useful for realization of a new source of entangled photons based
on the diamond optomechanical coupling with the dual-rail encoding. Such a
source could generate entangled photons at wavelengths inconvenient to
produce by other methods. For instance, we may generate entanglement between
ultraviolet and infrared photons, with the infrared photon good for quantum
communication and the ultraviolet photon convenient to be interfaced with
other qubits, such as the ion matter qubits. Such a photon source is hard to
generate by the conventional spontaneous parametric down conversion method.
In future, the tools based on the ultrafast pump and probe could be combined
with the powerful laser cooling or low-temperature technology, to provide
more efficient ways for quantum control of the optomechanical systems, with
important applications for realization of transduction of quantum signals
\textbf{\textbf{\cite{23,24}}}, processing of quantum information or
single-photon signals \textbf{\textbf{\cite{19,20,22a1}}}, and sensing of
small mechanical vibrations \textbf{\textbf{\cite{19}}}.

\section{Methods}

\subsection{Quantum process tomography}

Quantum process tomography (QPT) \cite{28} is defined by a completely
positive map $\varepsilon :\rho _{\text{f}}\equiv \varepsilon (\rho _{\text{i%
}})$ that transfers an arbitrary input state $\rho _{\text{i}}$ to the
output $\rho _{\text{f}}$. It can be characterized by a unique process
matrix $\chi _{mn}$ through the map $\rho _{\text{f}}=\sum_{mn}E_{m}\rho _{%
\text{i}}E_{n}^{\dagger }\chi _{mn}$ by choosing a fixed set of basis
operator ${E}_{m}$. In our experiment, we set the basis operators ${E}_{m}$
to be the identity operator $I$ and the three Pauli matrices $X=\sigma _{x}$%
, $Y=-i\sigma _{y}$, $Z=\sigma _{z}$. This corresponds to a choice of six
complementary input states $|H\rangle $, $|V\rangle $, $|+\rangle $, $%
|-\rangle $, $|L\rangle $, $|R\rangle $ for the teleportation. We
reconstruct the output state from teleportation by quantum state tomography
and use them to calculate the process matrix $\chi $ through the maximally
likelihood estimation \cite{28}. The process fidelity is determined by $F_{%
\text{p}}=Tr(\chi \chi _{\text{id}})$, where $\chi _{\text{id}}$ is the
identity process matrix corresponding to the perfect case. The process
fidelity $F_{\text{p}}$ determines the average teleportation fidelity $%
\overline{F}$ by the formula $\overline{F}=(2F_{\text{P}}+1)/3$ \cite{28},
where $\overline{F}$ is defined as the fidelity averaged over all possible
states of the input qubit with equal weight.

\textbf{Acknowledgements} This work was supported by the Ministry of
Education of China through its grant to Tsinghua University. LMD
acknowledges in addition support from the IARPA program, the ARL, and the
AFOSR MURI program.

\textbf{Author Contributions} L.M.D. designed the experiment and supervised
the project. P.Y.H., Y.Y.H., X.X.Y., X.Y.C., C.Z., L.H. carried out the
experiment. L.M.D. and P.Y.H. wrote the manuscript.

\textbf{Author Information} The authors declare no competing financial
interests. Correspondence and requests for materials should be addressed to
L.M.D. (lmduan@umich.edu).


\begin{thebibliography}{99}
\bibitem{1} Bennett, C. H., et al. Teleporting an unknown quantum state via
dual classical and Einstein-Podolsky-Rosen channels. \textit{Phys. Rev. Lett.%
} \textbf{70}, 1895-1899 (1993).

\bibitem{2} Briegel, H-J., Dur, W., Cirac, J. I. \& Zoller, P. Quantum
repeaters: The role of imperfect local operations in quantum communication.
\textit{Phys. Rev. Lett.} \textbf{81}, 5932--5935 (1998).

\bibitem{3} Gottesman, D., and Chuang I. L. Demonstrating the viability of
universal quantum computation using teleportation and single-qubit
operations. \textit{Nature} \textbf{402}, 390-393 (1999).

\bibitem{4} Duan, L. M., Lukin, M. D., Cirac, J. I. \& Zoller, P.
Long-distance quantum communication with atomic ensembles and linear optics.
\textit{Nature} \textbf{414}, 413--418 (2001).

\bibitem{6} Bouwmeester, D. \textit{et al.} Experimental quantum
teleportation. \textit{Nature} \textbf{390}, 575-579 (1997).

\bibitem{7} Boschi, D., \textit{et al.} Experimental realization of
teleporting an unknown pure quantum state via dual classical and
Einstein-Podolsky-Rosen channels. \textit{Phys. Rev. Lett.} \textbf{80},
1121-1125 (1998).

\bibitem{8} Furusawa, A. \textit{et al.} Unconditional quantum
teleportation. \textit{Science} \textbf{282}, 706-709 (1998).

\bibitem{8b} S. Takeda \textit{et al.} Deterministic quantum teleportation
of photonic quantum bits by a hybrid technique. \textit{Nature} \textbf{500}%
, 315-318 (2013).

\bibitem{9} Riebe, M. et al. Deterministic quantum teleportation of atomic
qubits. \textit{Nature} \textbf{429}, 734--737 (2004).

\bibitem{10} Barrett, M. D. et al. Deterministic quantum teleportation with
atoms. \textit{Nature} \textbf{429}, 737--739 (2004).

\bibitem{11} Olmschenk, S. \textit{et al.} Quantum teleportation between
distant matter qubits. \textit{Science} \textbf{323}, 486-489 (2009).

\bibitem{12} Krauter, H. et al. Deterministic quantum teleportation between
distant atomic objects. \textit{Nature Phys.} \textbf{9}, 400--404 (2013).

\bibitem{13} Steffen, L. \textit{et al.} Deterministic quantum teleportation
with feed-forward in a solid state system, \textit{Nature} \textbf{500},
319--322 (2013).

\bibitem{14} Pfaff, W. \textit{et al.} Unconditional quantum teleportation
between distant solid-state quantum bits. \textit{Science} \textbf{345},
532-535 (2014).

\bibitem{15} Sherson, J. F. \textit{et al.} Quantum teleportation between
light and matter. \textit{Nature} \textbf{443}, 557-560 (2006).

\bibitem{16} Chen, Y.-A., \textit{et al.} Memory-built-in quantum
teleportation with photonic and atomic qubits. \textit{Nature Phys.} \textbf{%
4}, 103-107 (2008).

\bibitem{17} Gao, W. B. \textit{et al.} Quantum teleportation from a
propagating photon to a solid-state spin qubit. \textit{Nature Commun.}
\textbf{4}, 3744 (2013).

\bibitem{18} Bussieres, F. \textit{et al.} Quantum teleportation from a
telecom-wavelength photon to a solid-state quantum memory. \textit{Nature
Photon.} \textbf{8}, 775--778 (2014).

\bibitem{19} Aspelmeyer, M., Kippenberg, T. J., and Marquardt, F. Cavity
optomechanics. \textit{Rev. Mod. Phys.} \textbf{86}, 1391-1452 (2014).

\bibitem{20} Lee, K. C. \textit{et al.} Entangling macroscopic diamonds at
room temperature. \textit{Science} \textbf{334}, 1253-1256 (2011).

\bibitem{21} Lee, K. C. \textit{et al.} Macroscopic non-classical states and
terahertz quantum processing in room-temperature diamond. \textit{Nature
Photon.} \textbf{6}, 41-44 (2012).

\bibitem{22} England, D. G. \textit{et al.} Storage and retrieval of
THz-bandwidth single photons using a room-temperature diamond quantum
memory. \textit{Phys. Rev. Lett.} \textbf{114}, 053602 (2015).

\bibitem{22a1} Fisher, K. A. G. \textit{et al. }Frequency and bandwidth
conversion of single photons in a room-temperature diamond quantum memory.
\textit{Nature Commun.} \textbf{7}, 11200, 2016.

\bibitem{22a2} Bustard, P. J. \textit{et al. }Raman-induced slow-light delay
of THz-bandwidth pulses. \textit{Phys. Rev. A} \textbf{93}, 043810 (2016).


\bibitem{23} Stannigel, K. Rabl, P., Soensen, A. S., Zoller, P. and Lukin,
M. D. Optomechanical Transducers for Long-Distance Quantum Communication.
\textit{Phys. Rev. Lett.} \textbf{105}, 220501 (2010).

\bibitem{24} Rabl, P. \textit{et al.} A quantum spin transducer based on
nanoelectromechanical resonator arrays. \textit{Nature Phys.} \textbf{6},
602--608 (2010).

\bibitem{24a} Dong, C.-H., Fiore, V., Kuzyk, M. C. and Wang, H.-L.
Optomechanical dark mode. \textit{Science} \textbf{338}, 1609-1613 (2012).

\bibitem{25} Gigan, S. \textit{et al.} Self-cooling of a micromirror by
radiation pressure. \textit{Nature} \textbf{444}, 67-70 (2006).

\bibitem{26} O'Connell, A. D. \textit{et al.} Quantum ground state and
single-phonon control of a mechanical resonator. \textit{Nature} \textbf{464}%
, 697-703 (2010).

\bibitem{27} Li, T.-C., Kheifets, S., Raizen, M. G. Millikelvin cooling of
an optically trapped microsphere in vacuum. \textit{Nature Phys.} \textbf{7}%
, 527--530 (2011).

\bibitem{28} White, A. G. \textit{et al.} Measuring two-qubit gates. \textit{%
J. Opt. Soc. Am. B} \textbf{24}, 172-183 (2007).

\bibitem{29} Blinov, B. B. \textit{et al.} Observation of entanglement
between a single trapped atom and a single ion \textit{Nature} \textbf{428},
153-157 (2004).
\end{thebibliography}
\end{document}